\documentclass[aps,prb,amsmath,twocolumn,showpacs]{revtex4}
\usepackage{graphicx}
\usepackage{dcolumn}% Align table columns on decimal point
\usepackage{bm}% bold math

\begin{document}
\title{Correlated electron-phonon transport from molecular dynamics with quantum baths}
\author{J. T. L\"u}
\email{tower.lu@gmail.com}
\author{Jian-Sheng Wang}
%\homepage[]{http://staff.science.nus.edu.sg/~phywjs/}
\affiliation{
Center for Computational Science and Engineering and Department of Physics,
National University of Singapore, Singapore 117542, Republic of Singapore
}
\date{31 May 2008}
\begin{abstract}
Based on generalized quantum Langevin equations for the tight-binding
wave function amplitudes and lattice displacements, electron and
phonon quantum transport are obtained exactly using molecular dynamics
(MD) in the ballistic regime. The electron-phonon interactions can be
handled with a quasi-classical approximation.  Both charge and energy
transport and their interplay can be studied.  We compare the MD
results with that of a fully quantum mechanical nonequilibrium Green's
function (NEGF) approach for the electron currents. We find a
ballistic to diffusive transition of the electron conduction in one
dimensional chains as the chain length increases.
\end{abstract}
\pacs{05.60.Gg, 72.10.Bg, 63.20.kd, 73.63.-b}
\maketitle

\section{Introduction}
The interaction of electrons with phonons in open nonequilibrium
molecular structures is of great importance within the context of
molecular electronics \cite{agrait,mgalperin07-mtj}. A variety of
methods at different levels of sophistication has been used to study
this problem, each working at a specific parameter range
\cite{mgalperin07-mtj,mitra2004}. The perturbative approach with a
self-consistent Born approximation (SCBA) works well when the
electron-phonon interaction (EPI) is weak, and has been used in the
first-principles study \cite{frederiksen:256601}. In the strong
interaction limit, it is possible to eliminate the bilinear EPI term
via a canonical transformation \cite{mahan-book}. This latter approach
has only limited use in a minimum model calculation, where there is
only one single electron degree of freedom (DOF) interacting with one
single phonon DOF. It is also possible to study the coherent
electron-phonon dynamics in the full coupling regimes using the
scattering theory \cite{nessfisher}, but this kind of methods ignores
dephasing between electrons and phonons. Hybrid approaches exist,
where the electron part is treated quantum-mechanically, while the
phonon system is handled by classical MD \cite{verdozzi:046603} with
quantum corrections \cite{horsfield05-cei}. Most of the above methods
are developed within the context of electronic transport. The
inclusion of phonon transport appears only very recently, mainly using
the NEGF approach \cite{galperin:155312,lu:165418}.

Molecular dynamics is usually viewed as a method that produces only
classical results.  In this paper, we introduce a new MD method to
study the correlated electron and phonon transport in open molecular
junctions for the quantum systems. It is based on a generalized
Langevin equation \cite{ford:504} for electrons and phonons, which so
far have been used to study their quantum transport separately
\cite{segal:6840,dhar-elangevin}. The formalism is exact in the
ballistic case, i.e., without the EPI. Quasi-classical approximation
\cite{schmid} is made to the full quantum many-body problem for
interacting systems.  It does not have to assume a bilinear form of
the EPI Hamiltonian, and it is applicable to the full electron-phonon
coupling range.  More importantly, the method can simulate large
systems. In the rest of the paper, we introduce a model system, derive
the quantum Langevin equations, and analyze the approximation
involved.  We present the MD numerical results of molecular chains,
and compare with those from NEGF method.

\section{Model and Theory}
Consider a typical $LCR$ structure for transport study, where a
molecular structure ($C$) is connected with two semi-infinite leads
($L$ and $R$) as electron and phonon reservoirs. The two leads are
linear systems in their respective thermal equilibrium states
characterized by the chemical potential and temperature. Possible
manybody interactions only exist in the central region.  The total
Hamiltonian is the sum of the two subsystems and their interaction,
$H_{\rm{e}} + H_{\rm{ph}} + H_{\rm{epi}}$.  The phonon part is
\begin{equation}
	H_{\rm{ph}} =\!\! \sum_{\alpha=L,C,R}\!\!\! H_{\rm{ph}}^\alpha \!+\! (u^L)^T V_{\rm{ph}}^{LC}u^C \!+\! (u^C)^T V_{\rm{ph}}^{CR} u^R \!+\! V_n,
	\label{eq:ph1}
\end{equation}
where $H_{\rm{ph}}^\alpha =
\frac{1}{2}(\dot{u}^\alpha)^T\dot{u}^\alpha + \frac{1}{2}(u^\alpha)^T
K^\alpha u^\alpha$. $u^\alpha$ is a column vector consisting of all
the displacement operators in the $\alpha$ region, and
$\dot{u}^\alpha$ is its conjugate momentum. The atomic mass has been
absorbed into $u_j = \sqrt{m_j}\,x_j$. $K^\alpha$ is the spring
constant matrix.  $V_{\rm{ph}}^{LC}$ is the coupling matrix between
the left lead and the central molecule, and
$V_{\rm{ph}}^{CL}=(V_{\rm{ph}}^{LC})^T$, similarly for
$V_{\rm{ph}}^{CR}$. $V_n$ is an anharmonic potential, which only
depends on $u^C$. The electron subsystem is given in a tight-binding
form in an orthogonal basis,
\begin{equation}
	H_{\rm e} = \!\! \sum_{\alpha=L,C,R}\!\!\! {c^\alpha}^\dagger T^\alpha c^\alpha  + 
\sum_{\alpha=L,R}\left({c^C}^\dagger V_{\rm e}^{C\alpha} c^\alpha
+\rm{h.c.}\right),
	\label{eq:e3}
\end{equation}
$c^\alpha$ (${c^\alpha}^\dagger$) is the column (row) vector
containing all the annihilation (creation) operators in the $\alpha$
region.  $V_{\rm e}^{C\alpha}$ has a similar meaning as $V_{\rm
ph}^{C\alpha}$, and $V_{\rm e}^{C\alpha} = \left(V_{\rm e}^{\alpha
C}\right)^\dagger$. h.c. represents Hermitian conjugate. The total
electron energy under the Born-Oppenheimer approximation depends on
the position of the atoms, so that we can make a Taylor expansion of
it about the atomic equilibrium positions, and obtain the
electron-phonon interaction terms (e.g., from a first-principles
calculation)
\begin{equation}
	H_{\rm epi} = \sum_{ijk}c^\dagger_i M^k_{ij} c_j u_k + \frac{1}{2}\sum_{i,j,k,l}c^\dagger_i M^{kl}_{ij} c_j  u_k u_l 
	 + \cdots,
	\label{eq:e2}
\end{equation}
$H_{\rm epi}$ includes all the higher order terms of the Taylor
expansion. The superscript $C$ has been omitted since EPI only takes
place in the center part.  $M^k_{ij}$ and $M^{kl}_{ij}$ are the first
and second order EPI coefficients, respectively.

Working in the Heisenberg picture, we obtain the equations of motion
for operators $u^\alpha$ and $c^\alpha$, e.g., for $c$,
\begin{eqnarray}
	i\, \dot{c}^\alpha &=& T^\alpha c^\alpha + V_{\rm e}^{\alpha C} c^C, \quad (\alpha=L, R),
	\label{eq:eom1} \\
	i\, \dot{c}^C &=& T^C c^C + V_{\rm e}^{CL} c^L + V_{\rm e}^{CR} c^R + [c^C,H_{\rm epi}].
	\label{eq:eom2}
\end{eqnarray}
We set $\hbar=1$, $e=1$ throughout the formulas. The lead operators can be
solved formally,
\begin{equation}
	c^\alpha (t) = ig_\alpha^r(t,t_1)c^\alpha (t_1) + \int_{t_1}^t g_\alpha^r(t,t')V_e^{\alpha C}  c^C(t') dt',
	\label{eq:eom3}
\end{equation}
where
$g^r_\alpha(t,t')=-i\theta(t-t')\langle[c^\alpha(t),{c^\dagger}^\alpha(t')]_+\rangle$
is the electron retarded Green's function for the lead $\alpha$.  It satisfies
\begin{equation}
	i\frac{\partial}{\partial t'}g_\alpha^r(t,t') + g_\alpha^r(t,t')T^\alpha  = -I\delta(t-t'),
	\label{eq:eom4}
\end{equation}
with the boundary condition $g^r_\alpha(t,t')=0~(t<t')$.  Using
Eq.~(\ref{eq:eom3}), the equation of motion of the central operator
reads
\begin{equation}
	i\dot{c}^C = T^C c^C + \int_{t_1}^t\!\! \Sigma^r(t,t') c^C(t')dt' + \xi 
			 + \sum_k M^ku_kc^C.
	\label{eq:eomf}
\end{equation}
Similar equation can be derived for the phonon displacement operators \cite{wang-prl07},
\begin{eqnarray}
	\ddot{u}^C = &-K^C u^C+F_n-\int_{t_1}^t \Pi^r(t,t')u^C(t')dt'+\eta \nonumber\\
			& -{c^{C}}^\dagger M c^C. 
	\label{eq:eom5}
\end{eqnarray} 
$F_n$ is the force due to anharmonic effect. The last terms of
Eqs. (\ref{eq:eomf}) and (\ref{eq:eom5}) are due to EPI. We have only
kept the first order term of the Taylor expansion, although inclusion
of higher orders is straightforward. Equations (\ref{eq:eomf}) and
(\ref{eq:eom5}) have the form of the generalized Langevin equation for
the quantum Brownian motion \cite{hanggi:026105}.

Let us try to understand these two equations. The damping kernels
$\Sigma^r=\Sigma_L^r+\Sigma_R^r$ and $\Pi^r=\Pi^r_L+\Pi^r_R$ are the
electron and phonon retarded self-energies in the NEGF formalism. They
are defined as, e.g., for electron
\begin{equation}
	\Sigma^r_{\alpha}(t,t') = V_{\rm e}^{C \alpha}g_{\alpha}^r (t,t') V_{\rm e}^{\alpha C},\quad (\alpha=L, R).
	\label{eq:se}
\end{equation}
In the wide-band limit, the coupling with the leads does not depend on
the energy. The damping kernel approaches memoryless $\delta$-function
in the time domain.  $\xi=\xi_L(t)+\xi_R(t)$ and
$\eta=\eta_L(t)+\eta_R(t)$ are electron and phonon random noises due
to the leads ($\alpha=L,R$)
\begin{equation}
\xi_\alpha(t) = iV_{\rm e}^{C\alpha}g^r_\alpha(t,t_1)c^\alpha(t_1), 
	\label{eq:en}
\end{equation}
and 
\begin{equation}
\eta_\alpha(t) = V_{\rm ph}^{C\alpha}\left[d^r_\alpha(t,t_1)\dot{u}^\alpha(t_1)-\dot{d}^r_\alpha(t,t_1)u^\alpha(t_1)\right].
	\label{eq:ph}
\end{equation}
$d^r_\alpha(t,t_1)=-i\theta(t-t_1)\langle[u^\alpha(t),u^\alpha(t_1)^T]\rangle$
is the lead retarded Green's function for phonons. In the leads the
electron and phonon subsystems do not couple. They are both linear
systems. In addition, the left and right lead are completely
independent.  The statistical properties of the random noises are
determined by the equilibrium ensembles at the remote pass, $t_1$.
Working in the eigenmode representation, we can show that the
expectation value of each noise term is zero. We can also obtain their
correlation matrices, e.g., for electrons
\begin{equation}
	\Xi^\alpha(t,t') = \langle \xi_\alpha^\dagger(t') \xi_\alpha^T(t)\rangle^{T} = -i\Sigma_\alpha^<(t-t').
	\label{eq:cor1}
\end{equation}
As expected, it does not depend on the initial time $t_1$, and is time
translationally invariant.  It is convenient to work in the Fourier
domain,
\begin{equation}
	\tilde{\Xi}^\alpha[\omega]= 
\int_{-\infty}^{+\infty}\!\!\!\!\Xi^\alpha(t-t')\, 
e^{i\omega (t-t')}dt = f^\alpha_{\rm e}(\omega) \Gamma_{\rm e}^\alpha[\omega].
	\label{eq:fourier}
\end{equation}
$f^\alpha_{\rm e}(\omega)$ is the Fermi distribution
function. $\Gamma_{\rm e}^\alpha[\omega] =
i(\Sigma_\alpha^r[\omega]-\Sigma_\alpha^a[\omega])$ denotes the
coupling with the leads. $\tilde{\Xi}^\alpha[\omega]$ is positive
semi-definite, as required from a classical noise correlation.  The
phonon noise has a similar relation. A symmetric form is used here
\cite{wang-prl07}
\begin{eqnarray}
	\tilde{F}^\alpha [\omega] &=& \frac{1}{2}\int\limits^{+\infty}_{-\infty}\!\!\!\!\Big(
\bigl\langle  \eta_\alpha(t) \eta^T_\alpha(t')\bigr\rangle+
\bigl\langle  \eta_\alpha(t') \eta^T_\alpha(t)\bigr\rangle^T\Big)
	e^{i\omega(t-t')}dt\nonumber\\
	& = & \left(f^\alpha_{\rm ph}(\omega)+\frac{1}{2}\right)\Gamma_{\rm ph}^\alpha[\omega],
	\label{eq:cor2}
\end{eqnarray}
where $f^\alpha_{\rm ph}(\omega)$ is the Bose distribution for phonons, $\Gamma_{\rm ph}$ is
similar to $\Gamma_{\rm e}$. 

We notice that the noise Eqs.~(\ref{eq:en}) and (\ref{eq:ph}) contain
operators, which satisfy anti-commutation or commutation
relations. Electrons and phonons need different treatment.
Equations~(\ref{eq:cor1}) and (\ref{eq:fourier}) are only applicable
to electrons. To study the hole transport, we need to use the
correlation matrix $\langle \xi_\alpha(t)
\xi_\alpha^\dagger(t')\rangle$. For phonons a symmetrization is needed
to eliminate an imaginary part of the correlation.  In both cases the
relation between the damping and the noise term is a kind of
manifestation of the quantum fluctuation-dissipation theorem.

The electrical and energy current can be obtained from different
methods. We can use the current continuity condition. In the case of a
discrete Hamiltonian, the electrical current from cell $j-1$ to cell
$j$ is, with only the lowest EPI term included,
\begin{equation}
	I_j = -i\Bigl( c^\dagger_j T_{j,j-1} c_{j-1}+\sum_{k} c^\dagger_j M_{j,j-1}^k c_{j-1} u_k-{\rm h.c.}\Bigr). 
	\label{eq:cur}
\end{equation}
We can also get the current from each lead by studying the time
derivative of the electron number
\begin{equation}
	I_{\alpha} = -\frac{dN_\alpha}{dt} = -i({c^C}^\dagger B_\alpha - {\rm h.c.}),
	\label{eq:cur2}
\end{equation}
where $B_\alpha = V^{C\alpha}c^{\alpha} = \xi_\alpha + \int_{t_1}^t
\Sigma_\alpha^r(t,t')c^C(t')dt'$. In the same way, the electron energy
current is
\begin{equation}
	I_\alpha^E = -\frac{dH_\alpha}{dt} = -(B^\dagger{\dot{c}^C} + {\rm h.c.}).
	\label{eq:cur3}
\end{equation}

So far the formal quantum Langevin equations are in terms of
operators. To perform a MD simulation, we need to turn the operators
into numbers. This is achieved by taking their quantum mechanical
expectation values at the beginning of the dynamics. It is reasonable
to assume that the central region and the two leads are decoupled at
that time. The two baths assume canonical equilibrium distributions,
and the central region is in an arbitrary state denoted by the density
matrix $\rho^C$. The expectation value of any operator $A^C$ is
$\langle A^C \rangle = \rm{Tr}\{\rho^C A^C\}$. Taking the expectation
value of these operators, generating the noise series using their
correlations \cite{wang-prl07}, the operator Langevin equations are
turned into $c$-number equations. For products of operators, mean-field
type approximation is used, e.g., $\langle c u\rangle \approx
\langle c \rangle \langle u \rangle$. 
MD simulation can be done using
these two equations. The final result is the ensemble average over the
initial states. To evaluate the electrical current, the operators in
Eqs.~(\ref{eq:cur}--\ref{eq:cur3}) are replaced by the $c$-numbers got
from MD simulation, and also $c^\dagger$ replaced by $c^*$, which is
the complex conjugate of $c$. By doing this, we have taken the
classical approximation to the operators.

One may cast doubt that this approximation may be too inaccurate to
give reasonable results for the fermionic system for the electrical
current.  However, we can show rigorously that for the ballistic case
the classical Langevin dynamics with the appropriate noises gives
exactly the same result as that predicted by the NEGF method
\cite{meir-wingreen,caroli,dhar-elangevin}. To do this, we write the
Langevin equations in the frequency domain
\begin{eqnarray}
	c^C[\omega] &=& G_0^r[\omega]\left(\xi[\omega]+ \!\!\int\! M^k u_k[\omega']c^C[\omega\!-\!\omega']\frac{d\omega'}{2\pi}\right),
	\label{eq:co} \\
	u^C[\omega] &=& D_0^r[\omega]\Bigg(-\eta[\omega]-F_n[\omega]+
\nonumber \\
&& \qquad \int {c^C}^\dagger[\omega]Mc^C[\omega-\omega']\frac{d\omega'}{2\pi}\Bigg).
	\label{eq:uo}
\end{eqnarray}
We also have
\begin{equation}
	B_\alpha[\omega] = \xi_\alpha[\omega]+\Sigma^r_\alpha[\omega]c^C[\omega].
	\label{eq:bo}
\end{equation}
In the ballistic case, we write Eq.~(\ref{eq:cur2}) in the energy
domain and substitute Eqs.~(\ref{eq:co}--\ref{eq:bo}) into it. Using
the noise correlation Eq.~(\ref{eq:fourier}), after some
rearrangement, we get exactly the Meir-Wingreen formula in the NEGF
method. In the presence of EPI, Eqs.~(\ref{eq:co}--\ref{eq:uo}) are
coupled. Repeated iteration with respective to $c^C[\omega]$ and
$u^C[\omega]$ gives an infinite series of terms. Analysis of these
terms shows that the quasi-classical approximation includes both the
crossed and the non-crossed Feynman diagrams. But it only reproduces
correctly part of the high order terms in the series, e.g., out of the
seven lowerest order nonlinear self-energy graphs, two of the graphs
involving $G^{>}$ is replaced by $-G^{<}$.  These wrong terms are not
important when the electron number per site in the center region is
small or the EPI is not strong, which defines the application range of
the quasi-classical approximation.

\section{Numerical Results and Discussions}
To illustrate the present approach, we take a simple one-dimensional
(1D) atomic chain connected with two 1D leads and simulate the coupled
equations (\ref{eq:eomf}) and (\ref{eq:eom5}) on computer. Each atom
has only one displacement degree of freedom and one spinless electron
state. We take the two leads to be the same with spring constant
$k_l$, hopping matrix element $-h_l$, and electron onsite energy
$\varepsilon_l$. $k_c$, $-h_c$, and $\varepsilon_c$ denote those of
the central part. Their couplings are $-v_e$ and $-v_{ph}$ for
electrons and phonons.  Some of the matrices, e.g., $T^C$ and
$V^{LC}_e$, are given by
\begin{equation}
T^C = \left( \begin{array}{cccc}
       \varepsilon_c & -h_c & 0 & \cdots \\
       -h_c & \varepsilon_c & -h_c & \cdots \\
       0 & -h_c & \varepsilon_c & -h_c \\
       \cdots & 0 & -h_c & \varepsilon_c 
       \end{array}
       \right), 
\end{equation}
\begin{equation}
V^{LC}_e = \left( \begin{array}{cccc}
      0 & \cdots &   & \\
      0 & 0 & \cdots   & \\
      -v_e & 0 & 0  & \cdots \\
       \end{array}
       \right). 
\end{equation}
The lead Green's functions have analytical solutions
\cite{lu:165418}. The anharmonic force $F_n$ is turned off in order to
perform a comparison with the NEGF method. The voltage is applied by
shifting the chemical potentials of the two leads.  A tight-binding
SSH type EPI term \cite{ssh}
\begin{equation}
	H_{\rm epi} = m \sum_{i = 1}^{L-1} \bigl(c_i^\dagger c_{i+1} + 
        c_{i+1}^\dagger c_i \bigr)(u_{i+1} - u_i)
	\label{eq:ssh}
\end{equation}
is used in the simulation. 
%%with
%%$M^{i}_{i,i+1}=M^{i}_{i+1,i}=-M^{i}_{i,i-1}=-M^{i}_{i-1,i}=-m$ and all others
%%are zero. 
The Langevin equations, with all the operators replaced by their
expectation values, are numerically solved using a fourth order
Runge-Kutta method.  A time-step of $\Delta t=5\times10^{-17}\,$s and
$10^6$ MD steps are used for each data point. As for the NEGF results,
the Meir-Wingreen expression for electrical current
\cite{meir-wingreen}, $I_\alpha = \frac{e}{2\pi}\int {\rm
Tr}\{G^>\Sigma^<_\alpha-G^<\Sigma^>_\alpha\}d\omega$, is used. The
greater (lesser) self-energy $\Sigma^>_\alpha[\omega]$
($\Sigma^<_\alpha[\omega]$) is due to the lead $\alpha$. $G^>[\omega]$
($G^<[\omega]$) is the greater (lesser) Green's functions of the
central region. A finite difference is used to calculate the quantum
conductance from the electrical current.

\begin{figure}
\includegraphics[scale=1.0]{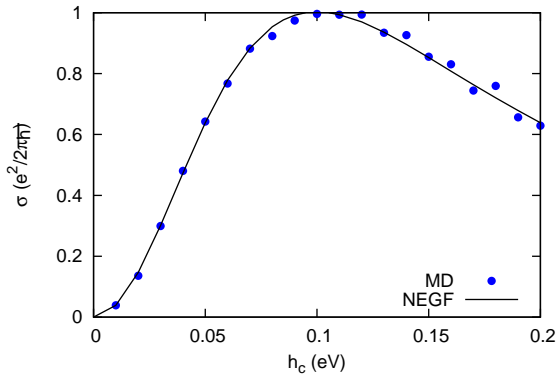}
\caption{\label{fig:1}Ballistic electron quantum conductance as a
function of hopping matrix element between the two atoms $h_c$ at $1$
K. Other parameters are $\varepsilon_c=\varepsilon_l=0$, $h_l=0.1$ eV,
$v_e=0.1$ eV. The line is from NEGF and the dots are MD.}
\end{figure}

\begin{figure}
\includegraphics[scale=1.0]{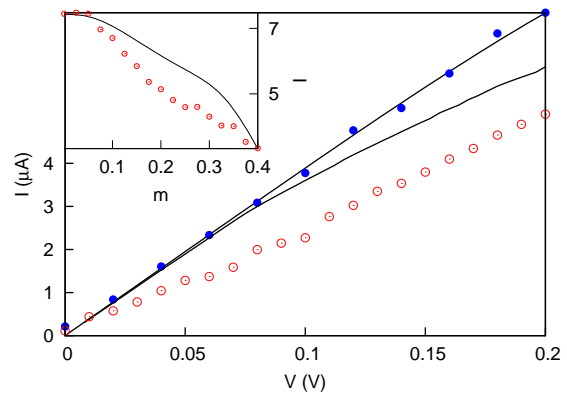}
\caption{\label{fig:2}Current-voltage characteristics of the two-atom
chain at $1$ K with the following parameters: $h_l=1.0$ eV, $h_c=0.1$
eV, $v_e=0.32$ eV, $\varepsilon_c=\varepsilon_l=0$, $k_l=k_c=0.5$
eV/(amu \AA$^2$), $v_{ph}=0.1$ eV/(amu \AA$^2$), and $m=0.2$
eV/(amu$^{\frac{1}{2}}$\AA). A small onsite spring constant
$k_0=0.2k_c$ is applied for the whole structure. MD results are shown
in points, and NEGF in lines. The filled dots and the straight line
are the ballistic results. The lower line
and the unfilled dots are results with EPI. The inset shows the
electrical current as a function of EPI strength $m$ at $V=0.2$ V.}
\end{figure}

We first demonstrate that the MD and the NEGF method give the same
results in the ballistic case. Figure~\ref{fig:1} shows the ballistic
electron conductance of a two-atom chain as a function of the hopping
matrix element between them $h_c$. When $h_c=0.1$ eV, the conductance
reaches a maximum value corresponding to one quantum unit
($e^2/2\pi\hbar$). Going apart from this value in both directions
leads to conductance decrease. The MD and the NEGF method give exactly
the same results within the statistical errors of the MD
simulation. This can also be seen from the ballistic $I$--$V$ curve in
Fig.~\ref{fig:2} (the upper curve in the main panel).

Now we turn on the EPI. The main panel of Fig.~\ref{fig:2} shows the
$I$--$V$ characteristics of the two-atom junction. The lower and upper
curves are with and without EPI, respectively. In the presence of EPI,
both methods give approximate results.  The NEGF results are based on
the SCBA, where only the non-crossed Feynman diagrams are included in
the self-energies \cite{lu:165418}. The MD method is non-perturbative
and includes both crossed and non-crossed diagrams, but only part of
these diagrams are treated correctly. As a result, the electrical
current from MD is lower than that from the NEGF method. This can also
be seen from the inset of Fig.~\ref{fig:2} where we change the
electron-phonon interaction streghth at an applied bias of $0.2$ V.

\begin{figure}
\includegraphics[scale=1.0]{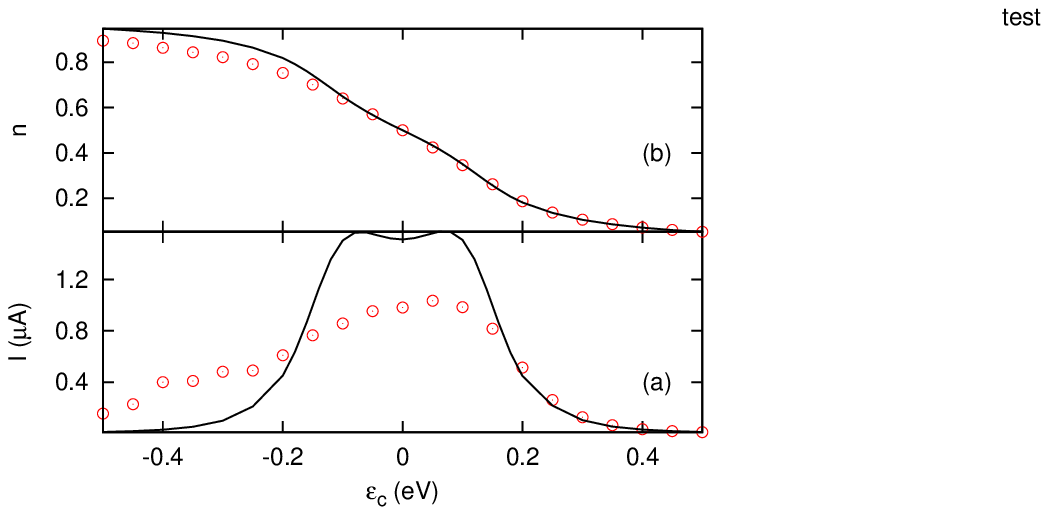}
\caption{\label{fig:3}The electrical current $I$ (a) and the
average electron number per site $n$ (b) at $V=0.04$ V as a
function of electron onsite energy $\varepsilon_c$ in the center. The
electron-phonon interaction $m=0.2$ eV/(amu$^{\frac{1}{2}}$\AA). All
other parameters are the same with Fig.~\ref{fig:2}.}
\end{figure}

A detailed analysis of the high order terms in the quasi-classical
approximation shows that its accuracy depends much on the electron
average occupation number in the center region. When the electron
number in small, the diagrams that the quasi-classical approximation
treats incorrectly is not important. In this regime, the MD method
should be accurate quantitatively. Out of this regime, it can only
gives qualitatively results.  These analysis is confirmed in
Fig.~\ref{fig:3}, where we show the electrical current and average
electron number per atom as a function of the electron onsite energy
in the center region. The electron number from the two methods shows
slight discrepancy only when the onsite energy is very low.  The MD
electrical current agrees with the NEGF method only when the electron
number is below $0.3$.

\begin{figure}
\includegraphics[scale=1.0]{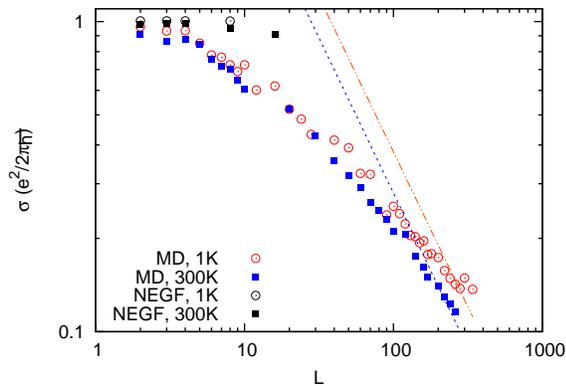}
\caption{\label{fig:4}Log scale plot of the electron conductance as a
function of chain length for $m=0.05$ eV/(amu$^{\frac{1}{2}}$\AA),
$h_l=h_c=v_e=0.1$ eV, and $\varepsilon_c=\varepsilon_l=0$.  Phonon
parameters are the same with Fig.~\ref{fig:2}.}
\end{figure}

The MD approach has its advantage: it can handle much larger systems
than the NEGF method. This is easy to understand. Given the total
degrees of freedom $N$, we only need to solve a set of $2N$ coupled
equations in the MD method. While in the NEGF method matrix
multiplication and inverse need much longer computer time (of order
$N^3$). In Fig.~\ref{fig:4}, we show the length dependence of the
electron conductance. Study of this effect using the NEGF method is
formidable. Besides the long computer time needed, convergence is also
hard to achieve for long chains. From the log scale plot, we find a
length independent conductance for short chains and close to inverse
linear ($1/L$) dependence for long chains. This corresponds to a
ballistic to diffusive transition of the electronic transport.  This
transition takes place earlier at $300$ K due to more available
phonons for scattering. Previous study of this transition relies on a
phenomenological method \cite{datta-book}. Thus a first-principle method that is able to
cover both regions is highly desirable. The MD method proposed here
could be one candidate.

The NEGF results should be valid at small values of EPI. But at
intermediate interaction range, no good approximation exists. From
this point of view, the MD method proposed here provides an
alternative nonperturbative way to study the correlated
electron-phonon dynamics in the intermediate EPI regime, although it
is only quantitatively accurate when the electron occupation number is
small. The MD method does not depend on the forms of the EPI Hamiltonian and
the phonon anharmonic potential, though not exploited here. More
importantly, it can handle much larger systems than the NEGF method.
Further improvement of the results may be obtained by including
higher order quantum corrections
\cite{prezhdo:6557,horsfield05-cei}. 
\section{Conclusions}
In summary, we have proposed a MD method to study the correlated electron and
phonon transport in open nonequilibrium molecular structures. It is based on
the generalized quantum Langevin equations. The effects of the leads are
reflected in the Langevin equations as noises and damping terms, which satisfy
the quantum fluctuation-dissipation theorem.  Quantum effects of the leads are
taken into account properly at least for the electrical or energy current
calculation. The method gives exact results for both electrons and phonons in
the ballistic transport regime.  When there is EPI, it is a quasi-classical
approximation. The approximation is valid when the electron occupation number
in the center region is small. The method shows its advantages in treating
large systems, where fully quantum-mechanical study is formidable. We
illustrate this by studying the ballistic to diffusive transition of the
electrical conductance in 1D chains. Although only examples of electrical
currents are presented here, it has other applications. For example, we can
also study thermoelectric transport in molecular structures.

\section*{Acknowledgments}
The authors thank Per Hedeg\aa rd, Mads Brandbyge, Jian Wang, Lifa Zhang,
and Yong Xu for discussions. This work was supported in part by a Faculty
Research Grant (R-144-000-173-101/112) of National University of Singapore.

\bibliography{qmd}

\end{document}